\documentclass[conference]{IEEEtran}
\IEEEoverridecommandlockouts
\usepackage{cite}
\usepackage{amsmath,amssymb,amsfonts}
\usepackage{algorithmic}
\usepackage{graphicx}
\usepackage{textcomp}
\usepackage{xcolor}
\usepackage{multirow}
\usepackage{makecell}
\usepackage{booktabs}
\usepackage{tabularx}
\def\BibTeX{{\rm B\kern-.05em{\sc i\kern-.025em b}\kern-.08em
    T\kern-.1667em\lower.7ex\hbox{E}\kern-.125emX}}
\begin{document}

\title{MGFF-TDNN: A Multi-Granularity Feature Fusion TDNN Model with Depth-Wise Separable Module for Speaker Verification}

\author{
\IEEEauthorblockN{Ya Li}
\IEEEauthorblockA{\textit{College of Computer Science} \\
\textit{South-Central Minzu University}\\
Wuhan, China \\
leia@scuec.edu.cn}
\and
\IEEEauthorblockN{Bin Zhou}
\IEEEauthorblockA{\textit{College of Computer Science} \\
\textit{South-Central Minzu University}\\
Wuhan, China \\
binzhou@mail.scuec.edu.cn}
\and
\IEEEauthorblockN{Bo Hu}
\IEEEauthorblockA{\textit{Wuhan Dongxin Tongbang} \\
\textit{Information Technology Co., Ltd.}\\
Wuhan, China \\
hubo@etah-tech.com}
}
\maketitle

\begin{abstract}
In speaker verification, traditional models often emphasize modeling long-term contextual features to capture global speaker characteristics. However, this approach can neglect fine-grained voiceprint information, which contains highly discriminative features essential for robust speaker embeddings. This paper introduces a novel model architecture, termed MGFF-TDNN, based on multi-granularity feature fusion. The MGFF-TDNN leverages a two-dimensional depth-wise separable convolution module, enhanced with local feature modeling, as a front-end feature extractor to effectively capture time-frequency domain features. To achieve comprehensive multi-granularity feature fusion, we propose the M-TDNN structure, which integrates global contextual modeling with fine-grained feature extraction by combining time-delay neural networks and phoneme-level feature pooling. Experiments on the VoxCeleb dataset demonstrate that the MGFF-TDNN achieves outstanding performance in speaker verification while remaining efficient in terms of parameters and computational resources.
\end{abstract}

\begin{IEEEkeywords}
speaker verification, depth-wise separable convolution, multi-granularity feature fusion
\end{IEEEkeywords}

\section{Introduction}
Speaker Verification (SV) is a task aimed at verifying whether a given speaker originates from a registered source. The SV system extracts speaker characteristics from a segment of speech to compare different speaker identity information. The extracted speaker embeddings can also be utilized in downstream tasks, such as speaker diarization,  speech synthesis, and related areas in speech processing. Obviously, one of the great challenges of SV is how to extract distinguishable speaker features. In the early stages of traditional speaker verification tasks, the GMM-UBM model \cite{REYNOLDS2000gmm-ubm} was widely used. Inspired by joint factor analysis \cite{kenny2007joint}, some researchers proposed the i-vector \cite{dehak2011front} method to optimize the voiceprint feature space. Speaker verification systems based on i-vector and probabilistic linear discriminant analysis \cite{prince2007probabilistic} gradually became the mainstream approach \cite{garcia2011analysis, pan2017vector}.

With the rapid development of deep learning technology, SV has also entered the era of deep learning, and many models and methods based on deep learning have emerged. The d-vector \cite{variani2014deep, richardson2015deep} was the first to apply deep learning to SV, followed by the x-vector \cite{snyder2017deep, snyder2018x-vector}, which extracts fixed-dimensional speaker features from variable-length speech segments. Deep learning-based SV has gradually evolved into two mainstream architectures. One utilizes Time-Delay Neural Network (TDNN) \cite{snyder2018x-vector, Desplanques_2020ecapa, yao23branch} as the backbone network to extract speaker features, and the other employs Residual Network (ResNet) \cite{zeinali2019but, zhou2021resnext, chen23eresnet}.
\begin{figure}[th]
  \centering
\includegraphics[width=\linewidth]{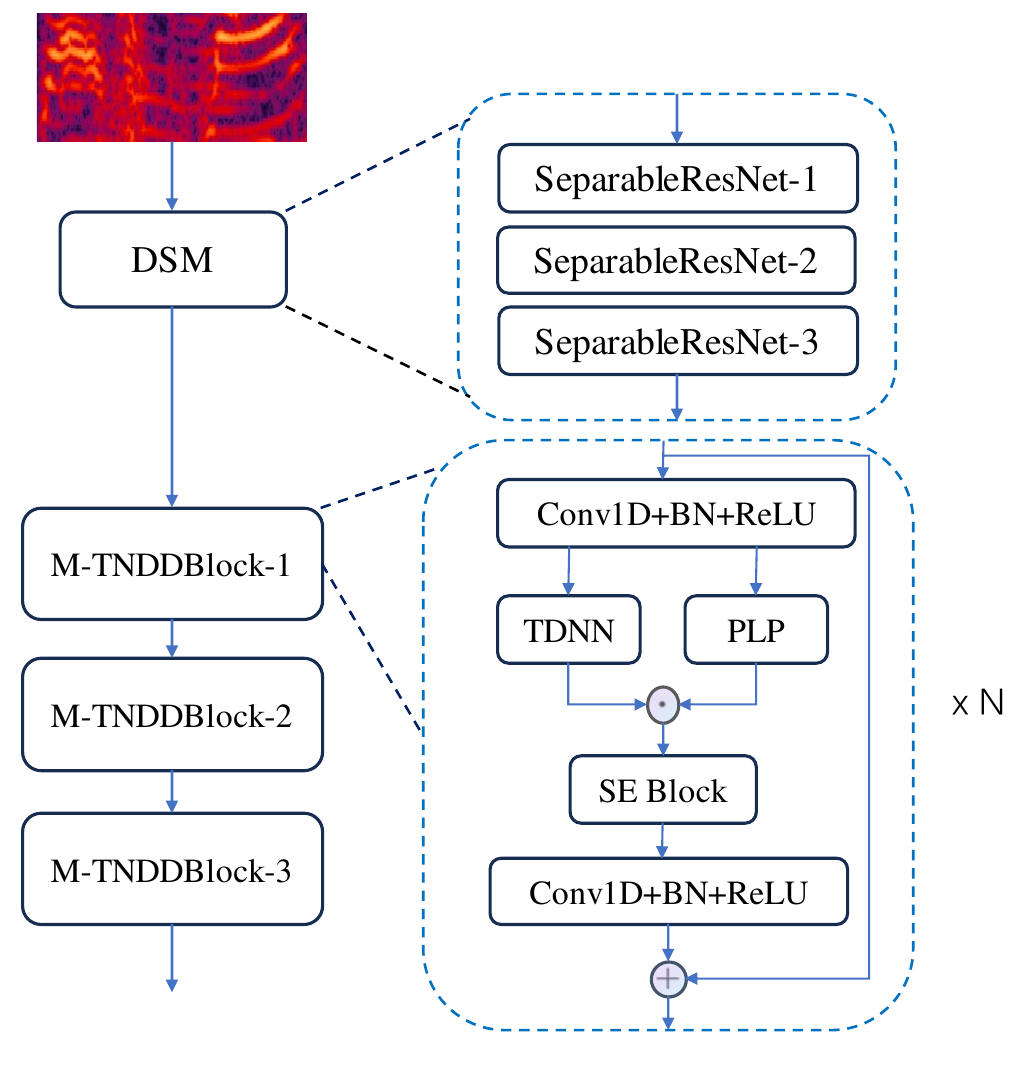}
  \caption{Overview of the proposed MGFF-TDNN architecture.}
  \label{fig:overview}
\end{figure}

Although TDNN-based models can effectively extract features across various temporal contexts, thoroughly capturing frequency-domain features often requires a substantial increase in model parameters. Similarly, ResNet-based models, though capable of modeling both temporal and frequency domains simultaneously, often struggle to achieve state-of-the-art (SOTA) performance and usually require a high number of parameters and significant computational resources. For instance, One successful variant model, ECAPA-TDNN \cite{Desplanques_2020ecapa}, has achieved SOTA performance, but at the expense of high parameters and computational complexity. ECAPA-TDNN employs the SE-Res2Block, which utilizes a one-dimensional Res2Net \cite{Gao_2021res2net} module combined with squeeze-excitation \cite{Hu_2018se} to extract frame-level features, achieving impressive performance. Models based on ResNet \cite{he2016resnet}, which use two-dimensional convolution to extract features in both the time and frequency dimensions, also demonstrate outstanding performance. However, these ResNet-based models often require a substantial number of parameters and computational resources to achieve competitive performance.

In this paper, we propose the MGFF-TDNN model, an enhanced TDNN-based architecture focused on multi-granularity context modeling, as illustrated in Fig.~\ref{fig:overview}. Firstly, to enhance the model's feature modeling in the frequency domain, a two-dimensional depthwise separable convolution module \cite{Sandler2018mobilenetv2} is employed as the front-end extractor to capture time-frequency domain features. Secondly, inspired by \cite{chen23eresnet, tan2022ponet}, we devise a Multi-granularity Temporal Delay Neural Network (M-TDNN) module. Within M-TDNN, a phoneme-level local max-pooling module is integrated to enhance fine-grained feature modeling. Additionally, the dilation factor of the TDNN module is progressively increased to extend the temporal context for feature modeling. Subsequently, the outputs of these two components are concatenated and fed into an Squeeze-Excitation (SE) block \cite{Hu_2018se} to capture internal dependencies among multi-granularity features. MGFF-TDNN is validated on the open-source VoxCeleb \cite{Nagrani17voxceleb,Chung18voxceleb2} dataset, and the experimental results demonstrate that the MGFF-TDNN architecture achieves competitive performance with fewer parameters and computational resources.

\section{System description}
\label{sec:3}
The overall architecture of the proposed MGFF-TDNN model is depicted in Fig.~\ref{fig:overview}. The architecture is primarily composed of two main blocks: the Depthwise Separable Module (DSM) and the Multi-Granularity Temporal Delay Neural Network Module (M-TDNN). Within the DSM, two-dimensional depth-wise separable residual network layers are introduced for the initial modeling of time-frequency features by extending the channel dimension. Subsequently, the generated feature maps are flattened along the channel and frequency dimensions and input into multiple M-TDNNBlocks, each of which contains a different number of M-TDNN layers. Each M-TDNN layer extracts features of varying granularities, which are then fused through the SE module to obtain attention weights for different granularity features.

\subsection{DSM block}
\begin{figure}[t]
  \centering
  \includegraphics[scale=0.5]{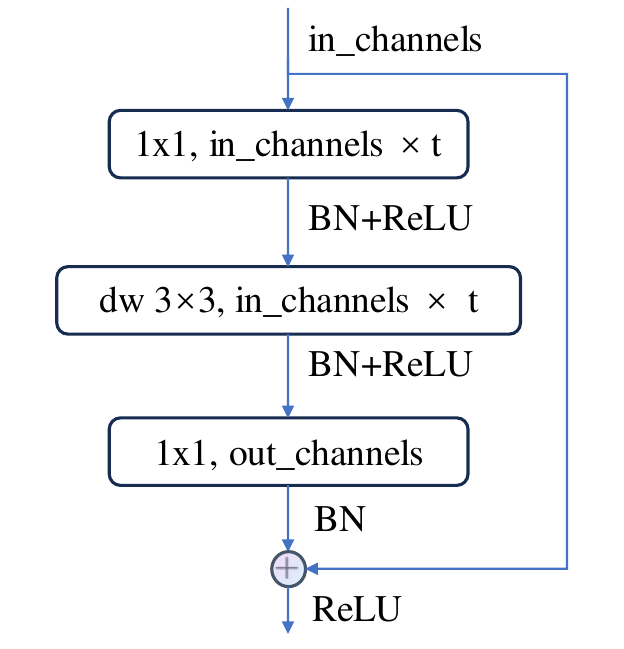}
  \caption{Illustration of depthwise separable residual block.}
  \label{fig:dwresnet}
\end{figure}

The vanilla time-delay neural network perform convolution operations along the temporal axis to capture time-domain features in the sequence. Although the convolutional kernels fully cover the frequency domain, a large number of filter banks are required to effectively model the frequency-domain features, which inadvertently increases the model's parameter count. To efficiently model both temporal and frequency-domain features, a two-dimensional convolution module is incorporated at the frontend. Inspired by \cite{Sandler2018mobilenetv2, 2022mfa}, we employ depthwise separable module in this paper, which comprises multiple layers of separable residual structures. Each layer's microstructure is referred to as an Inverted Residual Block \cite{Sandler2018mobilenetv2}. This structure was initially used in the field of computer vision, utilizing Depth-wise convolution paired with Point-wise convolution to extract features. This approach reduces the time and space complexity of convolutional layers while enhancing the network's representational capacity with minimal impact on performance. As illustrated in Fig.~\ref{fig:dwresnet}, assuming the input is x, the output y is obtained after three convolutional operations and residual connections. This process can be formally described by the following equation:
\begin{align}
\begin{split}
y = ReLU(x + BN( W_{3} \cdot ReLU( BN( W_{2} \cdot ReLU \\
(BN ( W_{1} \cdot x ) ) ) ) ) )
\end{split}
\end{align}
where $W_{1}$ and $W_3$ are the weights of the point-wise convolutions with output channels being $C*t$ and $C$ respectively, where t is the expansion factor typically set to $5 \sim 10$ (we set it to 6 in this work). $W_2$ is the weight of the depth-wise convolution, and its output channels are also $C*t$. $BN$ refers to batch normalization \cite{ioffe2015batch}, and $ReLU(\cdot)$ denotes rectified linear unit activation function. This module first employs point-wise convolution to increase the dimensionality of channels, then utilizes depth-wise convolution for feature extraction, and finally employs point-wise convolution for dimensionality reduction. This approach enables high-dimensional feature extraction, facilitating a comprehensive capture of the time-frequency domain characteristics in speech signals, while also strengthening the local modeling of acoustic features. Subsequently, the secondary point-wise convolution serves for dimensionality reduction, ensuring the maintenance of parameters and computational complexity within suitable bounds. Specific details are shown in Table \ref{tab:architecture}.

\begin{table}[th]
  %\centering
  \caption{Detailed architecture of the MGFF-TDNN network \hspace{0.5cm} Note: The MTDNN Block listed in the  table only enumerates the major variation units.}
    
    \resizebox{\columnwidth}{!}{%
    \renewcommand{\arraystretch}{1.4} % 调整整体行高
    
    \begin{tabular}{c|c|c}
    \toprule
    \textbf{Stage}      & \textbf{Structure}   & \textbf{Output\_size} \\
    \midrule
    \hline
    input & -     & 80×T \\
    \hline
    unsqueeze &   -    & 1×80×T \\
    \hline
    conv\_1 & 3×3,32 & 32×80×T \\
    \hline
    
    ResNet-1 &   $\begin{bmatrix}
                    {1 \times 1, 32\times 6} \\
                    {dw 3 \times 3,192,s=2} \\
                    {1 \times 1,32}
                    \end{bmatrix}$    & 32×40×T \\
    \hline
    ResNet-2 &   $\begin{bmatrix}
                    {1 \times 1, 32\times 6} \\
                    {dw3 \times 3,192,s=2} \\
                    {1 \times 1,32}
                    \end{bmatrix}$    & 32×20×T \\
    \hline
    ResNet-3 &   $\begin{bmatrix}
                    {1 \times 1, 32\times 6} \\
                    {dw3 \times 3,192,s=2} \\
                    {1 \times 1,32}
                    \end{bmatrix}$   & 32×10×T \\
    \hline
    reshape    &  -     & 320×T \\
    \hline
    MTDNN Block-1 &   $\begin{bmatrix}
                    {conv1d \times 1, 64} \\
                    {conv1d \times 3,64,d=1} \\
                    {conv1d \times 1,128}
                    \end{bmatrix} \times 3 $ & 128×T \\
    \hline
    MTDNN Block-2 &   $\begin{bmatrix}
                    {conv1d \times 1, 128} \\
                    {conv1d \times 3,128,d=2} \\
                    {conv1d \times 1,256}
                    \end{bmatrix} \times 6 $ & 256×T \\
    \hline
    MTDNN Block-3 &   $\begin{bmatrix}
                    {conv1d \times 1, 256} \\
                    {conv1d \times 3,256,d=2} \\
                    {conv1d \times 1,512}
                    \end{bmatrix} \times 4 $ & 512×T \\
    \hline
          & \multicolumn{2}{c}{Statistics pooling} \\
    \hline
          & \multicolumn{2}{c}{FC + BN} \\
    \hline
    \end{tabular}
    }
  \label{tab:architecture}%
\end{table}%

\subsection{M-TDNN block}
From an application perspective, a notable challenge in speaker verification tasks is modeling speakers from short utterances (typically only $3\sim5$ seconds). Unlike speaker modeling from longer utterances, which tends to focus more on contextual information, modeling from short utterances requires a finer attention to detail and more granular information. Under the same methodology, there is a higher probability of extracting rich information from longer utterances, leading to the neglect of the importance of granularity and instead emphasizing whether there is sufficient context. Experimental results from \cite{chen23eresnet} actually demonstrate the significant contribution of local feature fusion to model performance.

Based on the analysis above and inspired by \cite{tan2022ponet}, we propose a Multi-Granularity Feature Fusion (M-TDNN) module, which extracts features of different granularities and then employs specific fusion strategies to merge these features, thereby extracting important features of different granularities. Multi-granularity feature extraction differs from the previously mentioned multi-scale feature aggregation methods \cite{jung2020improving,gu2023memory}. Multi-scale methods focus more on integrating features across layers, combining contextual information of different lengths, while multi-granularity feature fusion emphasizes the integration of different granularity information within layers, avoiding a singular representation of features.

As shown in Fig.~\ref{fig:overview}, we denote the input of a layer in the M-TDNN Block as Y. Initially, Y undergoes one-dimensional convolution for preliminary feature extraction, resulting in the feature e:
\begin{gather}
  e = ReLU( BN( W_{1} \cdot Y ) )
\end{gather}
Subsequently, the feature e proceeds into two feature extraction branches. The left TDNN layer extracts dynamic contextual features $e_t$ by controlling the dilation factors, while the right phoneme-level pooling (PLP) extracts fine-grained features $e_p$ through overlapping sliding time windows.
\begin{align}
    e_{t} &= \tau(e) \\
    e_{p} &= plp(e)
\end{align}
Where $\tau(\cdot)$ represents the feature extraction process of the TDNN layer, and $plp(\cdot)$ represents phoneme-level pooling. Phoneme-Level Pooling (PLP) is implemented using standard max-pooling over sliding windows to capture fine-grained features. To reduce information loss, the PLP window slides with a 50\% overlap, and a sliding window size of 8 is utilized in this study. To ensure that the output tensor from PLP is aligned with the input tensor along the time dimension, we replicate the pooled features of the window across the time axis. This preparation facilitates subsequent feature fusion. We control the output dimensions of both TDNN and PLP, concatenate their outputs along the time axis, and then process them through the Squeeze-Excitation (SE) block to obtain $e_c$.
\begin{gather}
    s = \sigma( W_{2}\delta( W_{1} \cdot g \left \lbrack {e_{t},e_{p}} \right\rbrack + b_{1} ) + b_{2} ) \\
   e_{c} = \left\lbrack {e_{t},e_{p}} \right\rbrack \cdot s
\end{gather}
where $\sigma(\cdot)$ represents the sigmoid function, $\delta(\cdot)$ denotes the rectified linear activation function,  $[\cdot]$ signifies the feature concatenation process, and g refers to global pooling, which extracts global representations. Lastly, the weighted multi-granularity feature $e_c$ undergoes further feature fusion through one-dimensional convolution for enhanced feature extraction and is then residual-connected with the input $Y$, ultimately yielding the feature $e_o$.
\begin{align}
    e_{o} = ReLU( {Y + ReLU( {BN( {W_{1} \cdot e_{c}} )} )} )
\end{align}
The output $e_o$ serves as the output of the MTDNN layer, integrating features of different granularities. The detailed configurations are illustrated in Table \ref{tab:architecture}, where each MTDNN module consists of varying numbers of MTDNN layers. The three MTDNN modules include [3, 6, 4] MTDNN layers with output channels of 128, 256, and 512 respectively. The MTDNN modules in Table \ref{tab:architecture} outline the variation units within the MTDNN layers. To control the overall model parameters and computational complexity, we first employ one-dimensional convolution layers with a context of one frame to reduce the feature dimensions. This is followed by multi-granularity feature extraction. Subsequently, one-dimensional convolution is used for feature scaling and fusion, with the entire unit incorporating skip connections to enhance information flow.

\section{Experiments and analysis}
\label{sec:4}

\subsection{Datasets and evaluation metrics}
\label{section:metrics}

Experiments are conducted on the open-source speaker verification dataset, VoxCeleb \cite{Nagrani17voxceleb, Chung18voxceleb2}. Specifically, the development set of VoxCeleb2 \cite{Chung18voxceleb2} is utilized for training, which includes 5,994 speakers and a total of 1,092,009 speech segments. VoxCeleb1 \cite{Nagrani17voxceleb}'s development and test sets are employed for evaluation. The dataset includes three sets of trials with varying difficulty levels: VoxCeleb-O, VoxCeleb-H, and VoxCeleb-E. Given the complexity of acoustic environments, the training data is enhanced using noise datasets MUSAN \cite{snyder2015musan} and RIR \cite{ko2017rir}.

The trained models are evaluated using two common metrics: equal error rate (EER) and the minimum detection cost function (minDCF) with 0.01 target probability. 

\begin{table*}[ht]
  \centering
  \caption{EER and MinDCF(p-target=0.01) performance of different network architectures on the VoxCeleb1-O, VoxCeleb1-E, and VoxCeleb1-H test sets.}
    \begin{tabular}{cccccccc}
    \toprule
    \multirow{2}[4]{*}{Architecture } & \multirow{2}[4]{*}{\# Params(M) } & \multicolumn{2}{c}{VoxCeleb1-O} & \multicolumn{2}{c}{VoxCeleb1-E} & \multicolumn{2}{c}{VoxCeleb1-H}  \\
\cmidrule(lr){3-4} \cmidrule(lr){5-6} \cmidrule(lr){7-8}  &    & EER(\%) & MinDCF & EER(\%) & MinDCF & EER(\%) & MinDCF\\

    \midrule
    TDNN \cite{snyder2018x-vector}  & 4.62  & 2.31  & 0.322 & 2.37  & 0.273 & 4.25  & 0.393  \\
    ResNet18\cite{Desplanques_2020ecapa}& 13.8	&1.47&	0.177&	1.60&	0.179&	2.88&	0.267 \\
    Res2Net\cite{chen23eresnet}(Re-implemented)   & 4.03  & 1.37  & 0.139 & 1.27  & 0.141 & 2.26  & 0.216  \\
    ECAPA-TDNN \cite{Desplanques_2020ecapa}(Re-implemented) & 6.19 & 1.03  & 0.112 & 1.06  & 0.122 & 2.05  & 0.201  \\
    Branch-ECAPA-TDNN \cite{yao23branch} & 9.34 & 0.90  & 0.094 & 1.13  & 0.126 & 2.13  & 0.214  \\
    DS-TDNN \cite{ds-tdnn} & 6.50 & 0.90  & 0.118 & 1.15  & 0.140 & 2.11  & 0.199  \\
    MFA-TDNN \cite{2022mfa} & 5.93 & 0.97  & 0.091 & 1.14  & 0.121 & 2.17  & 0.199  \\
    D-TDNN \cite{yu2020densely} & 2.85  & 1.55  & 0.166 & 1.63  & 0.175 & 2.86  & 0.257  \\
    D-TDNN-L \cite{wang23cam++} & 6.40  & 1.19  & 0.118 & 1.21  & 0.129 & 2.22  & 0.205  \\
    \midrule
    MGFF-TDNN &  4.78  & \textbf{0.89}    & \textbf{0.086}      & \textbf{1.05}      & \textbf{0.119}      & \textbf{1.91}      &  \textbf{0.195} \\
    -w/o DSM &  4.81  & 1.14    & 0.117      & 1.17      & 0.128     & 2.11      &  0.206 \\
    -w/o PLP & 3.52	 & 1.04&	0.114&	1.08	&0.126	&2.00	&0.196 \\
    -w/o TDNN&	2.40&	1.22&	0.144&	1.38&	0.150&	2.47&	0.240 \\
    \bottomrule
    \end{tabular}%
  \label{tab:experiment}%
\end{table*}

\begin{table*}[htbp]
  \centering
  \caption{Performance comparison of different models on the VoxCeleb test set for both 3-second and 5-second duration groups.}
    \begin{tabular}{ccccccccccccc}
    \toprule
    \multirow{3}[3]{*}{Architecture} & \multicolumn{4}{c}{VoxCeleb1-O} & \multicolumn{4}{c}{VoxCeleb1-E} & \multicolumn{4}{c}{VoxCeleb1-H} \\
\cmidrule(lr){2-5}  \cmidrule(lr){6-9} \cmidrule(lr){10-13}          & \multicolumn{2}{c}{EER(\%)} & \multicolumn{2}{c}{MinDCF} & \multicolumn{2}{c}{EER(\%)} & \multicolumn{2}{c}{MinDCF} & \multicolumn{2}{c}{EER(\%)} & \multicolumn{2}{c}{MinDCF} \\
\cmidrule(lr){2-3} \cmidrule(lr){4-5} \cmidrule(lr){6-7} \cmidrule(lr){8-9} \cmidrule(lr){10-11} \cmidrule(lr){12-13}
          & 5s    & 3s    & 5s    & 3s    & 5s    & 3s    & 5s    & 3s    & 5s    & 3s    & 5s    & 3s \\
    \midrule
    Res2Net & 1.53  & 2.54  & 0.160 & 0.278 & 1.43  & 2.42  & 0.159 & 0.266 & 2.52  & 4.15  & 0.245 & 0.376 \\
    ECAPA-TDNN & 1.16  & 2.05  & 0.143 & 0.279 & 1.21  & 2.10  & 0.139 & \textbf{0.235} & 2.30  & 3.87  & \textbf{0.220} & 0.358 \\
    \midrule
    MGFF-TDNN & \textbf{1.04} & \textbf{1.93} & \textbf{0.129} & \textbf{0.215} & \textbf{1.19} & \textbf{2.06} & \textbf{0.133} & 0.236 & \textbf{2.18} & \textbf{3.65} & 0.221 & \textbf{0.357} \\
    \bottomrule
    \end{tabular}%
  \label{tab:short_dur}%
\end{table*}%

\begin{table}[htb]
  \centering
  \caption{The parameter counts, floating-point operations (FLOPs), and equal error rates (EER) on the VoxCeleb1-O test set for different models.}
    \begin{tabular}{cccc}
    \toprule
    \textbf{Network} & \textbf{Params(M)} & \textbf{FLOPs(G)} & \textbf{EER(\%)} \\
    \midrule
    Res2Net & 4.03  & 2.56 & 1.37 \\
    ECAPA-TDNN & 6.40  & 1.55 & 1.03 \\
    \textbf{MGFF-TDNN} & 4.78  & \textbf{1.49} & \textbf{0.89} \\
    \bottomrule
    \end{tabular}%
  \label{tab:complexity}%
\end{table}%

\subsection{Implementation details}
The proposed MGFF-TDNN model is implemented using the 3D-Speaker\footnote{https://github.com/alibaba-damo-academy/3D-Speaker} toolkit \cite{chen20243d}. For input features, we employ 80-dimensional log mel filterbank (FBank) features with a window length of 25ms and an offset of 10ms as input features. In addition to augmenting with noise datasets, speed perturbation is applied to the audio, with randomly sampling at rates of 0.9, 1.0, and 1.1 to triple the number of speakers. 

In our experiments, the stochastic gradient descent (SGD) optimizer is employed with an initial learning rate of 0.1, a momentum of 0.9, and weight decay set to 0.0001. We also incorporate a cosine annealing scheduler and linear warm-up scheduler for learning rate scheduling, with a minimum learning rate of 1e-4. The angular additive margin softmax (AAM-Softmax) loss \cite{deng2019arcface} is employed across all experiments, with margin and scaling factors set to 0.2 and 32, respectively. The dimension of the bottleneck in the SE-Block is set to 128. The final fully connected layer outputs speaker features of dimension 192. To enhance training efficiency, we randomly crop 3-second segments from each audio to construct training samples. During the evaluation phase, cosine similarity is used to calculate scores for computing the evaluation metrics mentioned in section~\ref{section:metrics}, which doesn't apply score normalization in the back-end.

\subsection{Results and analysis}
The experimental evaluation results comparing our proposed MGFF-TDNN to various baseline models are presented in Table \ref{tab:experiment}. From the table, it can be observed that our proposed method outperforms the baseline models. In comparison to the Res2Net model, our approach demonstrates significant improvements across all metrics. Specifically, it achieves relative improvements in EER of 35.0\%, 17.3\% and 15.5\%, and relative improvements in minDCF of 38.1\%, 15.6\% and 9.7\% on the three test sets, respectively. This limitation arises from the fact that Res2Net primarily focuses on modeling local time-frequency domain features, while lacking the capacity for global context modeling.

Compared to ECAPA-TDNN, our method achieves superior performance with fewer parameters across multiple test sets. Particularly on VoxCeleb1-O, the EER and minDCF improve by 13.6\% and 23.2\%, respectively. Although ECAPA-TDNN achieves multi-scale feature modeling in the temporal dimension, it lacks local modeling of time-frequency domain features, neglecting the impact of fine-grained feature modeling on the robustness of speaker embeddings. 

In the ablation study, the DSM module is removed, and only TDNN is used for simple dimensional mapping to ensure consistency with the input dimensions of the original model. The experimental results demonstrate the effectiveness of the DSM in extracting local time-frequency domain features. The performance degradation resulting from the removal of the PLP and TDNN modules highlights the significant role of multi-granularity feature fusion in advancing speaker representation modeling.

\begin{figure}[h]
  \centering
\includegraphics[width=\linewidth]{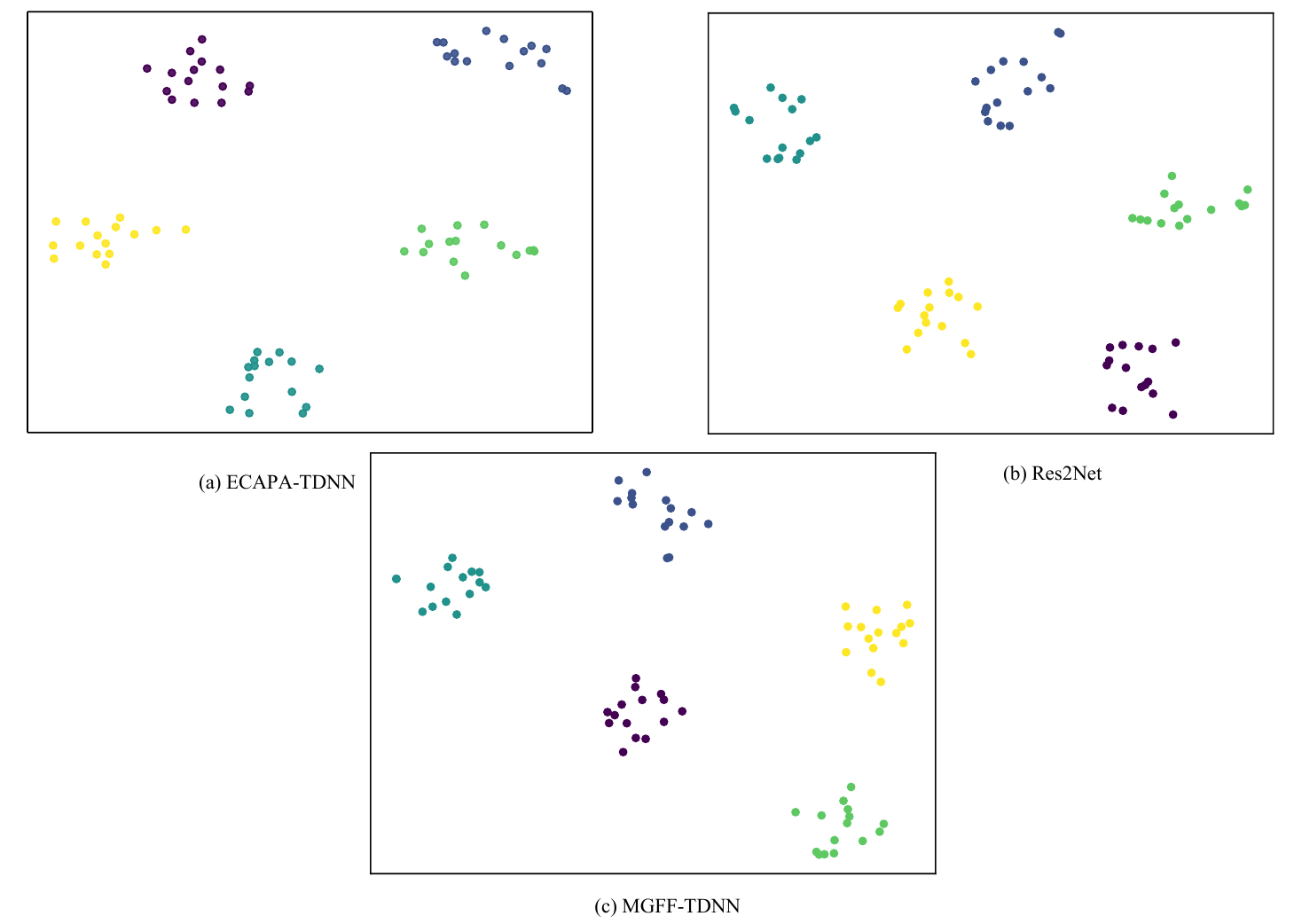}
  \caption{The t-SNE visualization depicts the extracted embeddings of five speakers. These 5-second speaker embeddings are derived from ECAPA-TDNN, Res2Net, and MGFF-TDNN models.}
  \label{fig:tsne}
\end{figure}

In order to show the proposed model's multi granularity feature fusion ability, we evaluate the performance of different models on the VoxCeleb test set for both 3-second and 5-second duration groups. As shown in Table \ref{tab:short_dur}, our proposed MGFF-TDNN model achieves superior performance on most metrics. Specifically, in the 3-second duration group, MGFF-TDNN shows EER reductions of 24.0\%, 14.9\%, and 12.0\% across the three test sets compared to Res2Net, and also demonstrates certain advantages over ECAPA-TDNN. Additionally, we visualize the 5-second speaker embeddings using t-distributed Stochastic Neighbor Embedding (t-SNE) \cite{tsne2008} and compare their disentanglement capabilities. As illustrated in Fig.\ref{fig:tsne}, the speaker embeddings extracted by MGFF-TDNN demonstrate stronger clustering capabilities in short duration compared to those from ECAPA-TDNN and Res2Net, and MGFF-TDNN makes speaker embeddings more discriminative. 

Firstly, the DSM module enhances the local modeling of time-frequency domain features. Building upon this, our proposed multi-granularity fusion module not only captures context information at different scales vertically through module stacking but also facilitates horizontal feature fusion across multiple granularities using diverse components. This dual approach addresses the challenge where single-granularity information may lead the model to neglect fine-grained features as the network deepens. Experimental results demonstrate that effective speaker embedding extraction requires attention to both time-frequency domain feature dimensions and the extraction of features at varying granularities across layers. Effective feature fusion methods play an important role in extracting robust speaker features. 

\subsection{Complexity analysis}
In this section, we analyze the computational complexities of the Res2Net, ECAPA-TDNN, and MGFF-TDNN models, focusing on their parameter counts and floating-point operations (FLOPs), as summarized in Table \ref{tab:complexity}. When comparing MGFF-TDNN to Res2Net, we observe that while MGFF-TDNN has a slightly higher parameter count, it achieves a significant reduction in FLOPs, indicating improved computational efficiency. In contrast, MGFF-TDNN outperforms both Res2Net and ECAPA-TDNN on the VoxCeleb1-O test set, despite exhibiting lower parameter counts and FLOPs than ECAPA-TDNN. This makes MGFF-TDNN a more efficient and effective model, especially in resource-constrained environments such as edge devices or situations with limited computational power, where lower FLOPs and memory requirements are crucial for real-time performance.

\section{Conclusion}
\label{sec:5}
In this paper, we propose a novel speaker verification model, termed MGFF-TDNN. Initially, two-dimensional depth-wise separable convolutions are employed for the pre-extraction of time-frequency domain features, facilitating the local modeling of both temporal and spectral information. Subsequently, multi-granularity feature fusion modules are introduced to capture features at different levels of granularity, thereby enabling the model to simultaneously learn fine-grained local patterns and broader contextual dependencies. This dual-level feature extraction strategy enhances the robustness of the speaker embeddings by integrating both localized and global contextual information. Experimental results on the VoxCeleb dataset demonstrate that MGFF-TDNN outperforms existing methods, achieving superior verification performance while maintaining a lower parameter count and reduced computational complexity.

\bibliographystyle{IEEEbib}
\bibliography{MGFF-TDNN}

\begin{thebibliography}{10}

\bibitem{REYNOLDS2000gmm-ubm}
Douglas~A. Reynolds, Thomas~F. Quatieri, and Robert~B. Dunn,
\newblock ``Speaker verification using adapted gaussian mixture models,''
\newblock {\em Digital Signal Processing}, vol. 10, no. 1, pp. 19--41, January 2000.

\bibitem{kenny2007joint}
Patrick Kenny, Gilles Boulianne, Pierre Ouellet, and Pierre Dumouchel,
\newblock ``Joint factor analysis versus eigenchannels in speaker recognition,''
\newblock {\em IEEE Transactions on Audio, Speech, and Language Processing}, vol. 15, no. 4, pp. 1435--1447, May 2007.

\bibitem{dehak2011front}
Najim Dehak, Patrick Kenny, Reda Dehak, Pierre Dumouchel, and Pierre Ouellet,
\newblock ``Front-end factor analysis for speaker verification,''
\newblock {\em IEEE Transactions on Audio, Speech, and Language Processing}, vol. 19, no. 4, pp. 788--798, May 2011.

\bibitem{prince2007probabilistic}
Simon~JD Prince and James~H Elder,
\newblock ``Probabilistic linear discriminant analysis for inferences about identity,''
\newblock in {\em 2007 IEEE 11th international conference on computer vision}, Rio de Janeiro, Brazil, October 2007, pp. 1--8.

\bibitem{garcia2011analysis}
Daniel Garcia-Romero and Carol~Y. Espy-Wilson,
\newblock ``{Analysis of i-vector length normalization in speaker recognition systems},''
\newblock in {\em Proc. Interspeech 2011}, Florence, Italy, August 2011, pp. 249--252.

\bibitem{pan2017vector}
Yilin Pan, Tieran Zheng, and Chen Chen,
\newblock ``I-vector kullback-leibler divisive normalization for plda speaker verification,''
\newblock in {\em 2017 IEEE Global Conference on Signal and Information Processing (GlobalSIP)}, Montreal, QC, Canada, November 2017, pp. 56--60.

\bibitem{variani2014deep}
Ehsan Variani, Xin Lei, Erik McDermott, Ignacio~Lopez Moreno, and Javier Gonzalez-Dominguez,
\newblock ``Deep neural networks for small footprint text-dependent speaker verification,''
\newblock in {\em 2014 IEEE international conference on acoustics, speech and signal processing (ICASSP)}, Florence, Italy, May 2014, pp. 4052--4056.

\bibitem{richardson2015deep}
Fred Richardson, Douglas Reynolds, and Najim Dehak,
\newblock ``Deep neural network approaches to speaker and language recognition,''
\newblock {\em IEEE signal processing letters}, vol. 22, no. 10, pp. 1671--1675, October 2015.

\bibitem{snyder2017deep}
David Snyder, Daniel Garcia-Romero, Daniel Povey, and Sanjeev Khudanpur,
\newblock ``Deep neural network embeddings for text-independent speaker verification,''
\newblock in {\em Proc. Interspeech 2017}, Stockholm, Sweden, August 2017, pp. 999--1003.

\bibitem{snyder2018x-vector}
David Snyder, Daniel Garcia-Romero, Gregory Sell, Daniel Povey, and Sanjeev Khudanpur,
\newblock ``X-vectors: Robust dnn embeddings for speaker recognition,''
\newblock in {\em 2018 IEEE international conference on acoustics, speech and signal processing (ICASSP)}, Calgary, AB, Canada, April 2018, pp. 5329--5333.

\bibitem{Desplanques_2020ecapa}
Brecht Desplanques, Jenthe Thienpondt, and Kris Demuynck,
\newblock ``{ECAPA-TDNN: Emphasized Channel Attention, Propagation and Aggregation in TDNN Based Speaker Verification},''
\newblock in {\em Proc. Interspeech 2020}, Virtual Event, Shanghai, China, October 2020, pp. 3830--3834.

\bibitem{yao23branch}
Jiadi Yao, Chengdong Liang, Zhendong Peng, Binbin Zhang, and Xiao-Lei Zhang,
\newblock ``{Branch-ECAPA-TDNN: A Parallel Branch Architecture to Capture Local and Global Features for Speaker Verification},''
\newblock in {\em Proc. Interspeech 2023}, Dublin, Ireland, August 2023, pp. 1943--1947.

\bibitem{zeinali2019but}
Hossein Zeinali, Shuai Wang, Anna Silnova, Pavel Matějka, and Oldřich Plchot,
\newblock ``But system description to voxceleb speaker recognition challenge 2019,''
\newblock {\em arXiv preprint arXiv:1910.12592}, 2019.

\bibitem{zhou2021resnext}
Tianyan Zhou, Yong Zhao, and Jian Wu,
\newblock ``Resnext and res2net structures for speaker verification,''
\newblock in {\em 2021 IEEE Spoken Language Technology Workshop (SLT)}, Shenzhen, China, January 2021, pp. 301--307.

\bibitem{chen23eresnet}
Yafeng Chen, Siqi Zheng, Hui Wang, Luyao Cheng, Qian Chen, and Jiajun Qi,
\newblock ``{An Enhanced Res2Net with Local and Global Feature Fusion for Speaker Verification},''
\newblock in {\em Proc. Interspeech 2023}, Dublin, Ireland, August 2023, pp. 2228--2232.

\bibitem{Gao_2021res2net}
Shang-Hua Gao, Ming-Ming Cheng, Kai Zhao, Xin-Yu Zhang, Ming-Hsuan Yang, and Philip Torr,
\newblock ``Res2net: A new multi-scale backbone architecture,''
\newblock {\em IEEE Transactions on Pattern Analysis and Machine Intelligence}, vol. 43, no. 2, pp. 652–662, February 2021.

\bibitem{Hu_2018se}
Jie Hu, Li~Shen, and Gang Sun,
\newblock ``Squeeze-and-excitation networks,''
\newblock in {\em Proceedings of the IEEE Conference on Computer Vision and Pattern Recognition (CVPR)}, Salt Lake City, UT, USA, June 2018, pp. 7132--7141.

\bibitem{he2016resnet}
Kaiming He, Xiangyu Zhang, Shaoqing Ren, and Jian Sun,
\newblock ``Deep residual learning for image recognition,''
\newblock in {\em Proceedings of the IEEE conference on computer vision and pattern recognition (CVPR)}, Las Vegas, NV, USA, June 2016, pp. 770--778.

\bibitem{Sandler2018mobilenetv2}
Mark Sandler, Andrew Howard, Menglong Zhu, Andrey Zhmoginov, and Liang-Chieh Chen,
\newblock ``Mobilenetv2: Inverted residuals and linear bottlenecks,''
\newblock in {\em Proceedings of the IEEE Conference on Computer Vision and Pattern Recognition (CVPR)}, Salt Lake City, UT, USA, June 2018, pp. 4510--4520.

\bibitem{tan2022ponet}
Chao-Hong Tan, Qian Chen, Wen Wang, Qinglin Zhang, Siqi Zheng, and Zhen-Hua Ling,
\newblock ``Ponet: Pooling network for efficient token mixing in long sequences,''
\newblock in {\em International Conference on Learning Representations (ICLR)}, Virtual Event, April 2022.

\bibitem{Nagrani17voxceleb}
Arsha Nagrani, Joon~Son Chung, and Andrew Zisserman,
\newblock ``{VoxCeleb: A Large-Scale Speaker Identification Dataset},''
\newblock in {\em Proc. Interspeech 2017}, Stockholm, Sweden, August 2017, pp. 2616--2620, ISCA.

\bibitem{Chung18voxceleb2}
Joon~Son Chung, Arsha Nagrani, and Andrew Zisserman,
\newblock ``{VoxCeleb2: Deep Speaker Recognition},''
\newblock in {\em Proc. Interspeech 2018}, Hyderabad, India, September 2018, pp. 1086--1090.

\bibitem{2022mfa}
Tianchi Liu, Rohan~Kumar Das, Kong Aik~Lee, and Haizhou Li,
\newblock ``{MFA: TDNN with Multi-Scale Frequency-Channel Attention for Text-Independent Speaker Verification with Short Utterances},''
\newblock in {\em 2022 IEEE International Conference on Acoustics, Speech and Signal Processing (ICASSP)}, Singapore, Singapore, May 2022, pp. 7517--7521.

\bibitem{ioffe2015batch}
Sergey Ioffe and Christian Szegedy,
\newblock ``Batch normalization: accelerating deep network training by reducing internal covariate shift,''
\newblock in {\em Proceedings of the 32nd International Conference on International Conference on Machine Learning - Volume 37}, Lille, France, July 2015, pp. 448--456.

\bibitem{jung2020improving}
Youngmoon Jung, Seong~Min Kye, Yeunju Choi, Myunghun Jung, and Hoi-Rin Kim,
\newblock ``{Improving Multi-Scale Aggregation Using Feature Pyramid Module for Robust Speaker Verification of Variable-Duration Utterances},''
\newblock in {\em Proc. Interspeech 2020}, Virtual Event, Shanghai, China, October 2020, ISCA, pp. 1501--1505.

\bibitem{gu2023memory}
Bin Gu, Wu~Guo, and Jie Zhang,
\newblock ``Memory storable network based feature aggregation for speaker representation learning,''
\newblock {\em IEEE/ACM Transactions on Audio, Speech, and Language Processing}, vol. 31, pp. 643--655, January 2023.

\bibitem{snyder2015musan}
David Snyder, Guoguo Chen, and Daniel Povey,
\newblock ``{MUSAN: A Music, Speech, and Noise Corpus},''
\newblock {\em arXiv preprint arXiv:1510.08484}, 2015.

\bibitem{ko2017rir}
Tom Ko, Vijayaditya Peddinti, Daniel Povey, Michael~L Seltzer, and Sanjeev Khudanpur,
\newblock ``A study on data augmentation of reverberant speech for robust speech recognition,''
\newblock in {\em 2017 IEEE International Conference on Acoustics, Speech and Signal Processing (ICASSP)}, New Orleans, LA, USA, March 2017, pp. 5220--5224.

\bibitem{ds-tdnn}
Yangfu Li, Jiapan Gan, Xiaodan Lin, Yingqiang Qiu, Hongjian Zhan, and Hui Tian,
\newblock ``Ds-tdnn: Dual-stream time-delay neural network with global-aware filter for speaker verification,''
\newblock {\em IEEE/ACM Transactions on Audio, Speech, and Language Processing}, vol. 32, pp. 2814--2827, 2024.

\bibitem{yu2020densely}
Ya-Qi Yu and Wu-Jun Li,
\newblock ``{Densely Connected Time Delay Neural Network for Speaker Verification},''
\newblock in {\em Proc. Interspeech 2020}, Virtual Event, Shanghai, China, October 2020, pp. 921--925.

\bibitem{wang23cam++}
Hui Wang, Siqi Zheng, Yafeng Chen, Luyao Cheng, and Qian Chen,
\newblock ``{CAM++: A Fast and Efficient Network for Speaker Verification Using Context-Aware Masking},''
\newblock in {\em Proc. Interspeech 2023}, Dublin, Ireland, August 2023, pp. 5301--5305.

\bibitem{chen20243d}
Yafeng Chen, Siqi Zheng, Hui Wang, Luyao Cheng, Tinglong Zhu, Changhe Song, et~al.,
\newblock ``3d-speaker-toolkit: An open source toolkit for multi-modal speaker verification and diarization,''
\newblock {\em arXiv preprint arXiv:2403.19971}, 2024.

\bibitem{deng2019arcface}
Jiankang Deng, Jia Guo, Jing Yang, Niannan Xue, Irene Kotsia, and Stefanos Zafeiriou,
\newblock ``Arcface: Additive angular margin loss for deep face recognition,''
\newblock {\em IEEE Transactions on Pattern Analysis and Machine Intelligence}, vol. 44, no. 10, pp. 5962--5979, October 2022.

\bibitem{tsne2008}
Laurens van~der Maaten and Geoffrey Hinton,
\newblock ``Visualizing data using t-sne,''
\newblock {\em Journal of Machine Learning Research}, vol. 9, no. 86, pp. 2579--2605, 2008.

\end{thebibliography}

\end{document}